\documentclass[twocolumn,preprintnumbers,amsmath,amssymb,prl]{revtex4}

\usepackage{epsfig}
\usepackage{graphicx}
\usepackage{dcolumn}
\usepackage{hyperref}
\usepackage{bm}

\begin{document}

\title{Temperature dependence of the slip length in polymer melts at attractive surfaces}

\author{J. Servantie}
\author{M. M{\"u}ller}
\affiliation{Institut f{\"u}r Theoretische Physik, Georg-August-Universit{\"a}t, 
37077 G{\"o}ttingen, Germany}

\date{\today}

\begin{abstract}
Using Couette and Poiseuille flow, we extract the temperature dependence of
the slip length, $\delta$, from molecular dynamics simulations of a
coarse-grained polymer model in contact with an attractive, corrugated
surface. $\delta$ is dictated by the ratio of bulk viscosity and surface
mobility. At weakly attractive surfaces, a lubrication layer forms, $\delta$ is large and increases upon cooling. Close to the glass transition temperature, $T_g$, very large slip lengths are observed. At a more attractive surface, a``sticky surface layer" is build up, which gives rise to a small slip length. Upon cooling, $\delta$ decreases at high temperatures, passes through a minimum and grows upon approaching $T_g$. At strongly attractive surfaces, the Navier-slip condition fails to describe Couette and Poiseuille flow simultaneously. The simulation results are corroborated by a schematic, two-layer model suggesting that the observations do not depend on the details of the computational model.
\end{abstract}

\maketitle

Rapid miniaturization of microfluidic devices has led to new questions in
hydrodynamics \cite{Quake2005} which often involve the role of
boundaries. The microscopic structure at the solid-fluid interface dictates
wettability and friction and much effort has been directed towards
tailoring these properties. The effect of microscopic surface properties
can be incorporated into a continuum description via a boundary condition (HBC)
to the Navier-Stokes equation. Theoretical and experimental studies of
fluid flow in confined systems have demonstrated that the no-slip boundary
condition, which postulates that the tangential velocity at the surface
vanishes, can be violated \cite{Tropea2007} and the fluid slips past the surface. In the context of microfluidics, slip is often preferred because it reduces the amount of shear stress or pressure drop required to maintain flow. Hence, higher through-put can be generated in microfluidic devices. The amount of slip can be quantified by the slip length, $\delta$. The slip boundary condition was formulated by Navier in 1823 as a balance between viscous stress and friction stress at the surface:
\begin{equation}
\left.\eta \frac{\partial v_x}{\partial z}\right|_{z=z_h}=\left. \frac{\eta}{\delta}
  v\right|_{z=z_h}
\label{EQ1}
\end{equation}
where $\lambda=\eta/\delta$ denotes the friction coefficient and $\eta$ the
bulk viscosity of the liquid. $z_h$ characterizes the position at which the
hydrodynamic boundary condition is to be applied. Both parameters, $\delta$
and $z_h$, characterize the HBC. The value of the slip length, $\delta$, has been measured experimentally for a variety of systems and the results have attracted abiding interest \cite{Tropea2007}. Typically, the scale of the slip length in simple liquids is set by the size of the constituents of the fluid \cite{Bocquet2006}. There are, however, notable exceptions: For instance, topographically structured surfaces have been designed to generate large slip, nano-bubbles or the formation of a lubrication layer at the solid-fluid interface due to surface segregation of a low-molecular weight component in a polydisperse melt can increase the slip length. Recently, a careful analysis of the rim profile of dewetting polymer films has also provided evidence of slip length of the order of micrometers, which is significantly larger than microscopic length scales in the system \cite{Jacobs2006}. 

In this Letter, we investigate the impact of surface interactions and the
concomitant effective viscosity \cite{Goel2008} at the surface on the HBC
by molecular dynamics simulations of a standard, coarse-grained polymer
model \cite{Grest1986}. We show that (i) the stronger is the attraction
between fluid particles and substrate the smaller is the slip length, (ii)
slip lengths can grow large as the glass transition temperature of the
fluid is approached, and (iii) for strongly attractive surfaces the Navier
slip condition (\ref{EQ1}) fails to provide a HBC that simultaneously
describes Poiseuille and Couette flow. Our simulation results are
corroborated by a phenomenological two-layer model.

In our model, polymers are comprised of $N=10$ coarse-grained segments that
interact via a truncated and shifted Lennard-Jones potential with cut-off,
$r_c=2 \sqrt[6]{2} \sigma$. Length and energy scales are set by the
parameters $\sigma$ and $\epsilon$ of the potential. Neighboring segments
are bonded together by a FENE (Finitely Extensible Non-linear Elastic)
potential \cite{Grest1986}. The model exhibits a glass transition
temperature and the dynamics in the bulk and in confinement has been
recently reviewed \cite{Baschnagel2005}. A Dissipative Particle Dynamics
(DPD) thermostat is used to control temperature, $T$. The solid surface is modeled by two rigid layers of
Lennard-Jones interaction centers arranged on a FCC lattice
\cite{Servantie2008}. By varying the strength, $\epsilon_s$, of the
Lennard-Jones potential between solid and fluid, we tune the adhesion and
slip. 

In order to mimic experimental conditions, the simulations are performed at
the coexistence pressure where the polymer melt coexists with a vapor of
vanishingly low density.  Rather than using a constant pressure algorithm
we tune the distance, $H$, between the
walls to attain the coexistence density. The system size is chosen large
enough for the properties in the center of the film to be independent from
the strength of the solid-fluid interaction. Periodic boundaries are applied in both, $x$ and $y$ directions and the dimensions of the simulation cell are $L_x=19.84\; \sigma$ and $L_y=19.93\; \sigma$, respectively. At low temperature the fluid consists of 20\,000 coarse-grained  segments and the distance between the walls varies between $51\;\sigma$ and $57\;\sigma$. At higher temperature $k_BT/\epsilon>0.8$, when the effect of the substrate propagates not that far into the film, we use 10\,000 particles, and $H$ varies between $30\;\sigma$ and $38\;\sigma$.

\begin{figure}
\includegraphics[scale=0.55]{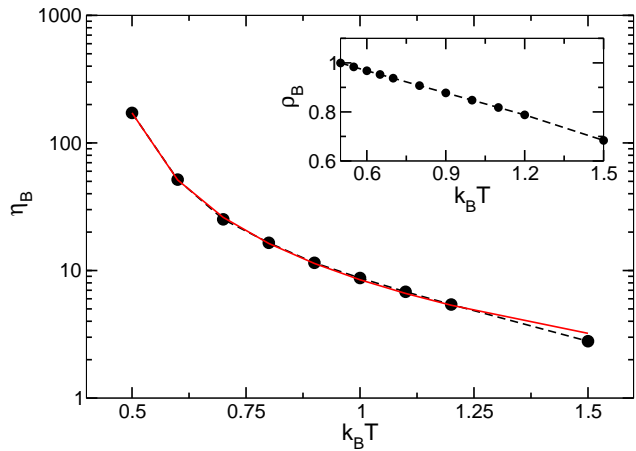}
\caption{(color online). Bulk viscosity as a function of temperature. The  circles mark the results of the molecular dynamics simulations and the solid line a fit using the power law predicted by mode-coupling theory. The inset presets the liquid-vapor co-existence density, $\rho_L$, versus temperature, $T$.}
\label{FIG1}
\end{figure} 

As a first result, we depict in Fig.~\ref{FIG1} the temperature dependence of the bulk viscosity extracted from the velocity profile, $v(z)$, of Poiseuille flow at the center of the film. The steady-state Navier-Stokes equation yields
\begin{equation}
\frac{\partial^2 v}{\partial z^2}=-\frac{\rho_L f}{\eta_B}
\label{EQNAVIER}
\end{equation}
where $\rho_L$, $\eta_B$, and $f$ denote the number density of the liquid,
the bulk viscosity, and the volume force applied to the segments,
respectively. The inset represents the liquid density, $\rho_L$, at
coexistence. Upon reducing temperature, the model exhibits a glass
transition at $T_g$. The bulk viscosity above the glass transition
temperature, $T_g$, is well describable by the power law, $\eta_B(T)
\propto |T-T_g|^{-\gamma}$ predicted by Mode Coupling theory
\cite{Gotze1992}. Fitting the simulation data in Fig.~\ref{FIG1} yields
$k_BT_g/\epsilon=0.41$ and $\gamma=1.57$, which agrees with previous studies \cite{Benneman1998}.  

Due to pronounced layering effects at the solid surface, the effective, near-surface
viscosity, $\eta_S$, deviates from the bulk behavior. While viscosity is
only properly defined in the bulk, there are several ways to estimate an
effective local viscosity. One consists in computing the local shear stress
and defining $\eta_s$ as the ratio between the local shear stress and the
velocity gradient. However, the gradient of the velocity profile is
difficult to extract with sufficient accuracy and in previous studies an
explicit form of the velocity profile has been assumed
\cite{Varnik2002}. Here, we instead analyze the local
mobility of the fluid. Assuming that the mobility can be estimated by the
Einstein relation $D \propto 1/\eta$, we obtain the estimate $\eta_B/\eta_S
\propto D_S/D_B$  
\footnote{For the motion of a coarse-grained segment in a viscous fluid the
  Einstein formula  $D = k_BT / 3 \pi \eta \sigma$ relates the
  self-diffusion coefficient $D$ to the viscosity of the surrounding
  fluid. For short chain lengths and low temperatures close to $T_g$, the
  fluid viscosity is dominated by non-bonded interactions and chain
  connectivity is less important. In fact, the Einstein relation describes
  the bulk behavior surprisingly well.}. 
 
\begin{figure}[t]
\includegraphics[scale=0.55]{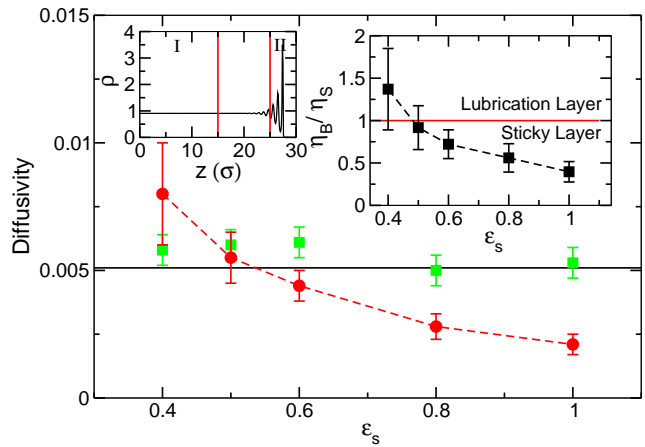}
\caption{(color online). Local diffusivity as a function of the strength of
  the substrate $\epsilon_s$ at $k_BT/\epsilon=0.8$. The circles and the
  squares are respectively the diffusivity near to the substrate and in the
  bulk. The straight line marks the Einstein estimate $D = k_BT / 3 \pi
  \eta \sigma$, for the fluid of unconnected segments. The right inset
  represents the effective viscosity ratio, $\eta_B/\eta_S$, as a function
  of $\epsilon_s$, the straight line delineates the formation of a lubrication layer and a sticky layer at the solid-fluid interface. The left inset depicts density profiles, and the vertical lines define the bulk and surface regions used in the calculations.}
\label{FIG2}
\end{figure}

The local diffusivity is obtained from the mean square displacement parallel to the surfaces as
$D_\alpha=\lim_{t\to \infty} \langle \Delta {\bf r}_{xy}^2 \rangle_\alpha/4t$
where only particles that remain in region $\alpha$ for the entire time of
the calculation contribution to the average \cite{daoulas2005}. We have
computed the local diffusivity at temperature $k_B T=0.8\epsilon$ for
$\epsilon_s=0.4,0.5,0.6,0.8,1.0\; \epsilon$. The boundaries for the bulk
region are chosen such that the bulk properties are independent of the
substrate strength. The fluid properties gradually change as a function of
the distance from the solid-liquid interface. Our choice of the width of
the boundary region is a compromise: It is wide enough such that segments
remain sufficiently long in the boundary region to determine the
diffusivity, and it is narrow enough in order for the properties to be dominated by surface effects. As we decrease $\epsilon_S$, the near-surface mobility becomes larger as shown in Fig.~\ref{FIG2} and the average time a particle stays in the surface region decreases. In the bulk region, the diffusivity is independent from the solid-fluid attraction, $\epsilon_S$.  In the inset, we depict the ratio $D_S/D_B$ and observe that substrate strengths larger than approximately $\epsilon_S=0.5\;\epsilon$ result in a ratio $\eta_B/\eta_S<1$, i.e., a ``sticky surface layer'' is formed. For $\epsilon_S < 0.5\;\epsilon$, the surface mobility is enhanced and a lubrication layer forms at the solid-fluid interface.

In order to measure the slip length, $\delta$, and position, $z_h$, of the HBC, we simultaneously compute Couette and Poiseuille flow profiles \cite{Pastorino2008,Servantie2008} for the different solid-fluid interaction strengths as a function of temperature. The shear rate employed is small enough such that $\delta$ does not depend on the strength of the flow \cite{Troian2004}. To simulate planar shear flow (Couette) the surfaces are moved at constant velocity, $v_S$. We typically employed $v_S=0.2 \sigma/\tau$ where $\tau$ denotes the reduced Lennard-Jones time unit. Poiseuille flow is generated by applying a force, $f$ on all particles. In our simulations, the volume force varies between $f=0.001-0.005 \epsilon/\sigma$. Complementary information is obtained by calculating the friction coefficient, $\lambda$, from equilibrium molecular dynamics by integrating the transverse force auto-correlation function \cite{Bocquet1994}. Previous studies for our model have shown that this procedure gives rise to consistent results at high temperatures \cite{Servantie2008}.

The velocity in the bulk region at the center of the film, $z=0$, are fitted by
linear or parabolic profiles as predicted by macroscopic hydrodynamics for
Couette and Poiseuille flow. Let $z_C$ and $z_P$ denote the positions,
where these linear and parabolic profiles extrapolate to vanishing
velocity, then the slip length and the position of the hydrodynamic boundary condition are given by:
\begin{equation}
\delta = \sqrt{z_C^2-z_P^2} \quad \mbox{and} \quad z_h=z_C+\delta
\label{EQSIMPLE}
\end{equation} 
\begin{figure}[b]
\includegraphics[scale=0.55]{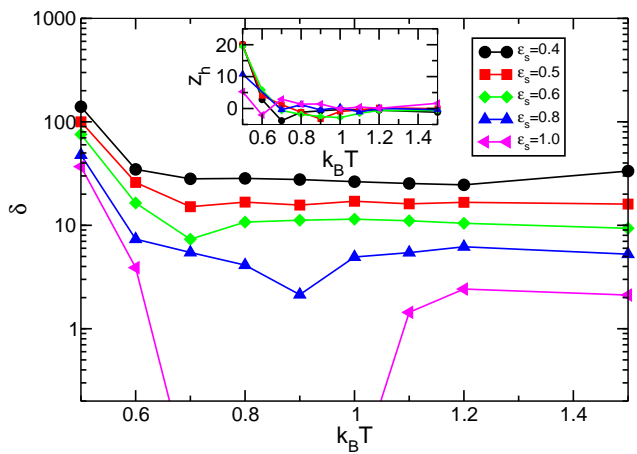}
\caption{(color online). Slip length, $\delta$, versus temperature for
  different strengths, $\epsilon_S$, of solid-fluid interaction. The inset
  represents the position of the hydrodynamic boundary $z_h$ versus temperature.}
\label{FIG3}
\end{figure}
The results for the slip length are depicted in Fig.~\ref{FIG3} and the
distance to the wall of the hydrodynamic boundary $z_h$ in the inset. We
see that the temperature dependence of $\delta$ is highly sensitive to
temperature and  solid-fluid interaction. First, we observe that
independently  from $\epsilon_s$ the slip length diverges as we approach
the glass transition temperature, $T_g$, because the fluid eventually behaves like a solid. Already at
$k_BT/\epsilon=0.5$, i.e., about $20\%$ above the glass transition
temperature of our model, $\delta$ has increased by an order of magnitude
compared to the approximately constant value at high temperature. This
observation offers an explanation for the surprisingly large slip length
observed in the dewetting experiment of Fetzer {\em et al.} \cite{Jacobs2006}, which were performed in the vicinity of the glass transition temperature.

The results for the position, $z_h$, of the HBC do not vary significantly with the strength of the fluid-solid interaction over the entire temperature regime. At high temperature, the position $z_h$ is close to the top of the solid surface, while it gradually shifts inwards as the temperature is reduced towards $T_g$. This effect goes along with a growing distance over which the liquid structure is altered by the surface as can be observed from the pronounced packing effects in the density profile. 

While the position, $z_h$, does not significantly depend on $\epsilon_S$, the behavior of the slip length qualitatively changes with the strength of the solid-fluid interaction. While the values of $\delta$ corresponding to $\epsilon_s=0.4\; \epsilon$ and $\epsilon_s=0.5\; \epsilon$ decrease monotonously with $T$, we observe a non-monotonous variation of the slip length for larger attraction, $\epsilon_S$. For very strong attraction, $\epsilon_S=1$, there is even a region of intermediate temperatures where $z_C<z_P$ and thus Eq.~(\ref{EQSIMPLE}) has no solution. This marks the failure of the Navier slip condition to parameterize the near-surface flow solely by the material properties of the solid surface. 

\begin{figure}[b]
\includegraphics[scale=0.33]{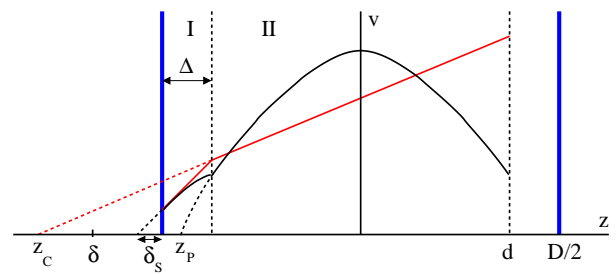}
\caption{(color online). Sketch of the Couette and Poiseuille flows in the schematic two-layer model}
\label{FIG4}
\end{figure}
In order to rationalize these simulation results and explore whether they are universal or specific to our coarse-grained polymer model, we propose a schematic two-layer model depicted in Fig.~\ref{FIG4}. Within this model, we approximate the gradual variation of the fluid properties as a function of the distance from the solid surface by a  boundary region of width, $\Delta$, which is characterized by a surface viscosity, $\eta_S$, and the bulk with viscosity, $\eta_B$. Within each layer, the fluid is described by the Navier-Stokes equation. At the interface between the solid surface and the boundary layer, we impose a Navier slip condition (\ref{EQ1}) with a microscopic slip length, $\delta_S$. At the interface between the surface layer and bulk, we require the continuity of shear stress and velocity,

\begin{equation}
\left.\eta_S \frac{\partial v_S}{\partial z} \right|_{z=-d}=\left.\eta_B  \frac{\partial v}{\partial z} \right|_{z=-d} 
,\quad
\left.v_S\right|_{z=-d}=\left.v\right|_{z=-d}
\label{EQBCOND}
\end{equation}

For planar shear flow one obtains for the linear velocity profile in the central bulk region
\begin{equation}
v=\frac{\Sigma}{\eta_B}\left[z+d\left(1-\frac{\eta_B}{\eta_S}\right)+\frac{\eta_B}{\eta_S}\left(\delta+z_h\right)\right]
\end{equation}
where $\Sigma$ quantifies the shear stress. If a volume force $f$ is applied to the fluid, one obtains a parabolic velocity profile in the boundary layer and in the bulk. Using Eq.~(\ref{EQNAVIER}) and the boundary conditions (\ref{EQ1}) and (\ref{EQBCOND}), we obtain for the latter:

\begin{equation}
v=\frac{\rho_L  f}{\eta_B}\left[-\frac{z^2}{2}+\frac{d^2}{2}\left(1-\frac{\eta_B}{\eta_S}\right)+\frac{1}{2}\frac{\eta_B}{\eta_S}\left(2z_h\delta+z_h^2\right)\right]
\end{equation}

Finally, Eq.~(\ref{EQSIMPLE}) yields the slip length
\begin{equation}
\delta=\sqrt{\Delta\frac{\eta_B}{\eta_S}\left(\frac{\eta_B}{\eta_S}-1\right)\left(\Delta+2\delta_S\right)+\left(\frac{\eta_B}{\eta_S}\delta_S\right)^2}
\label{EQDELTAA}
\end{equation}

The first term describes the effect of the surface layer, the second term arises from the microscopic slip at the solid surface. For surfaces with a large surface mobility, $\eta_B/\eta_S>1$, a lubrication layer is formed and results in an enhanced slip length, $\delta>\delta_S$, compared to the microscopic slip at the solid-fluid interface. On the other hand, if the solid-fluid interactions give rise to a boundary layer with large effective viscosity, $\eta_B/\eta_S<1$, the presence of this sticky layer at the substrate reduces the slip length, $\delta<\delta_S$). Moreover, if
\begin{equation}
\frac{\eta_B}{\eta_S}\le\frac{1+2\delta_S/\Delta}{\left(1+\delta_S/\Delta\right)^2}
\label{EQNOSLIP}
\end{equation}
the velocity far away from the surface cannot be described by the
Navier-Stokes equation and a Navier slip boundary condition (\ref{EQ1}). 

This schematic model can rationalize the observations in our molecular
dynamics simulation: (i) At high temperature, kinetic effects will dominate
the behavior, thus $\eta_S \approx \eta_B$. In this case, $\delta$ is equal
to the microscopic slip length $\delta \approx\delta_S$. (ii) Upon cooling
the fluid, the bulk viscosity increases. If the solid-fluid interactions
are weak, $\epsilon_S<0.5$, a lubrication layer is formed and the slip
length increases, $\delta \propto (\eta_B/\eta_S) \;\delta_S$. (iii) If the coupling between solid and fluid is strong, however, the ratio $\eta_B/\eta_S$ decreases upon cooling and so does $\delta$. If the ratio becomes sufficiently small (see Eq.~(\ref{EQNOSLIP})), as it does in the case $\epsilon_S=1$ for our model, the Navier slip condition fails.  Upon approaching $T_g$ from above, however, the slip length passes through a minimum and increases. The latter effect stems from the temperature dependence of the microscopic slip length, $\delta_S=\eta_S/\lambda$, which diverges for $T \to T_g$.
  
In conclusion, surface interactions can modify the near-surface mobility and thus can be exploited to control the hydrodynamic boundary condition. Very large slip lengths can be expected in the vicinity of the glass transition of the fluid. Depending on the strength of the interaction between the solid surface and the fluid, slippage may be enhanced or reduced, and at strongly attractive surfaces the Navier slip condition may even fail to provide an appropriate boundary condition to the Navier-Stokes equation with parameters that solely depend on the surface material. These findings of the simulations are corroborated by a schematic two-layer model which shows that the effects are universal and are not a consequence of the non-Newtonian nature of the polymer liquid employed in our study. 

We thank K. Ch. Daoulas, K. Jacobs, R. Seemann, C. Pastorino, and B. Wagner
for stimulating discussions. Financial support was provided by the DFG
priority program ``nano- and microfluidics" under grant Mu 1674/3. We gratefully acknowledge computer time at the J{\"u}lich Supercomputer Center. 

\bibliographystyle{apsrev}

\end{document}